\documentclass[12pt,preprint]{aastex}

\slugcomment{ApJ, in press}
\shorttitle{X-ray emission of SN 2004et with {\it Chandra} }
\shortauthors{}
\newcommand{\ergs}{\rm ~erg~s^{-1}}
\newcommand{\kms}{km~s$^{-1}$}

\newcommand{\chandra}{\textit{Chandra}}

\begin{document}
\title{{\it CHANDRA} OBSERVATIONS OF SN 2004et AND THE X-RAY EMISSION
OF TYPE IIP SUPERNOVAE}
\author{J. Rho\altaffilmark{1},  T. H. Jarrett\altaffilmark{1}, 
N. N. Chugai\altaffilmark{2,3}, R. A. Chevalier\altaffilmark{3}}

\altaffiltext{1}{{\it Spitzer} Science Center, California Institute of Technology, 
Pasadena, CA 91125; rho, jarrett@ipac.caltech.edu}
\altaffiltext{2}{Institute of Astronomy, RAS, Pyatnitskaya 48, 109017 Moscow, Russia;
nchugai@inasan.ru}
\altaffiltext{3}{Department of Astronomy, University of Virginia, P.O. Box 400325, 
Charlottesville, VA 22904; rac5x@virginia.edu}
\begin{abstract}

We report the X-ray detection  of the Type II-plateau supernova SN
2004et in the spiral galaxy NGC 6946, using the {\it Chandra X-Ray
Observatory}. 
The position of the X-ray source was found to agree with
the optical position within $\sim$0.4$''$. {\it Chandra} also surveyed
the region before the 2004 event, finding no X-ray emission at the
location of the progenitor. For the post-explosion observations, a total
of 202, 151, and 158 photons  were detected in three  pointings, each
~29 ks in length, on 2004 October 22, November 6, and December 3,
respectively. The spectrum of the first observation is best fit by a
thermal model  with a temperature of $kT=1.3^{+\infty}_{-0.8}$ keV and a
line-of-sight absorption of $N_H=1.0\pm1.0 \times 10^{22}$ cm$^{-2}$.
 The inferred unabsorbed luminosity ($0.4-8$
keV)  is $\sim$4$\times  10^{38}$ erg s$^{-1}$, adopting a  distance of
5.5 Mpc.  A comparison between hard and soft counts 
on the first and third  epochs indicates
a softening over this time,
although there is an insufficient number
of photons to constrain the variation of  temperature and absorption
by spectral fitting.
We model the emission as arising from the reverse shock
region in the interaction between the supernova ejecta and the
progenitor wind. For a Type IIP supernova with an extended progenitor,
the cool shell formed at the time of shock wave breakout from the star
can affect the initial evolution of the interaction shell and the
absorption of radiation from the reverse shock. The observed spectral
softening might be due to decreasing shell absorption. We find a
pre-supernova mass loss rate of $(2-2.5)\times 10^{-6}~M_{\odot}$
yr$^{-1}$ for a wind velocity of 10 \kms, which is in line with
expectations for a Type IIP supernova. 

\end{abstract}

\section{INTRODUCTION}

\begin{figure}
\includegraphics[width=15truecm,angle=0]{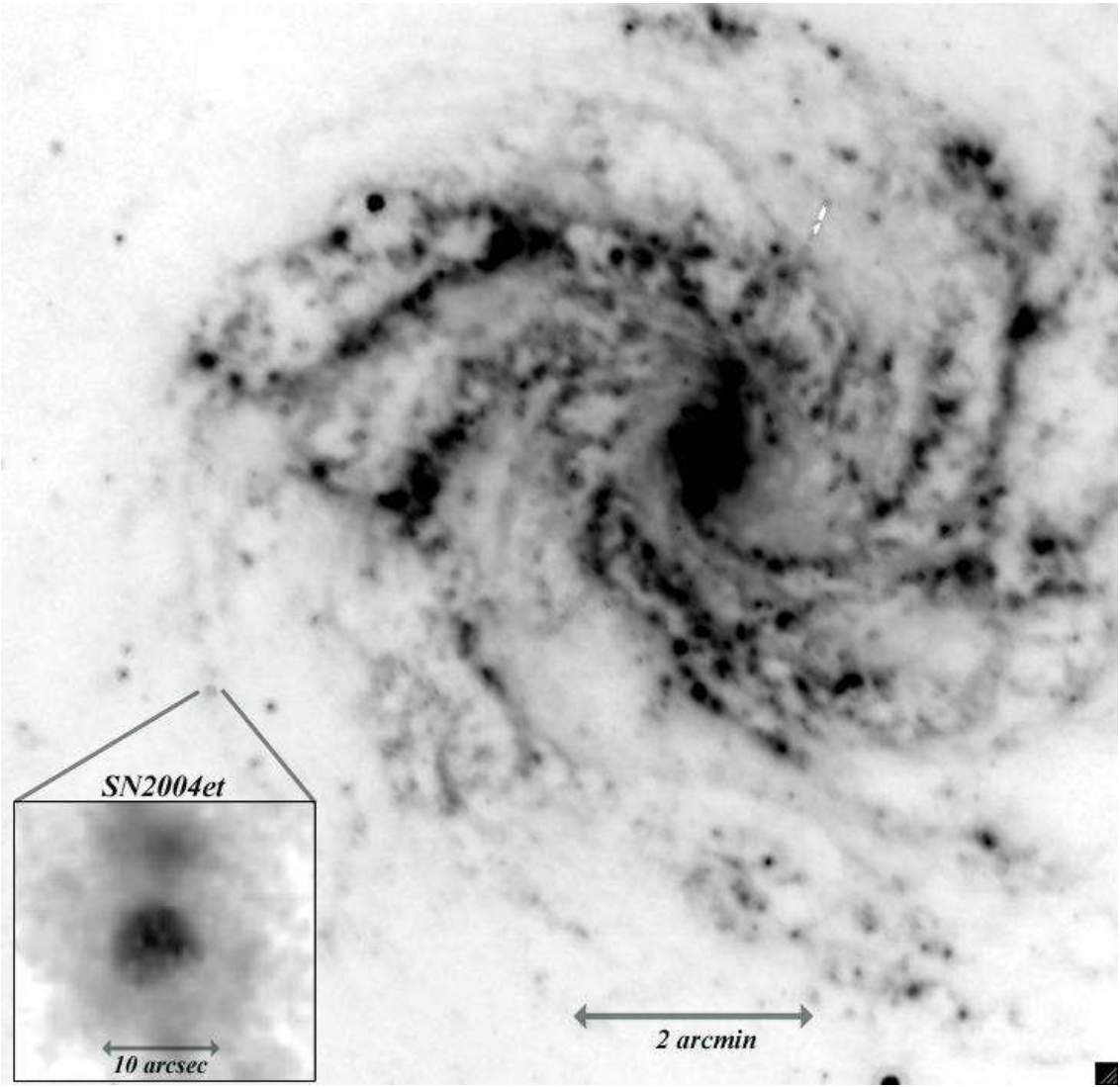}
\caption{
{\it Spitzer} IRAC 8 $\mu$m (from 6.5 to 9.5 $\mu$m) image  of NGC 6946
acquired by the SINGS \citep{ken03} project.
The observations were made on 2004 November 25, and the integration time was
480 sec per sky position. 
SN 2004et is located in the outer, south-eastern spiral arm;
a closeup view of the region is shown in the lower left, showing that
the source is extended, composed 
of a point source coinciding with SN 2004et and an extended shell 
that is $\sim$7 arcsec ($\sim$186 pc) in diameter. 
}
  \label{SN2004etspitzerimage}
\end{figure}

The supernova (SN) SN 2004et was  optically discovered on 2004 September
27 by \cite{zwi04}  at the position of R.A.\ $20^{\rm h}  35^{\rm m}
25.33^{\rm s}$ and Dec.\  $+60^\circ$07$^{\prime} 17.7^{\prime \prime}$
(J2000) in the spiral galaxy NGC 6946.  \cite{klo04} report detection of
the SN at magnitude $15.17\pm0.16$ on September 22.983 UT, 
$12.7\pm 0.3$ on September 25.978, and a non-detection
on September 22.017 with an lower magnitude limit of $19.4\pm 1.2$. 
The rapid rise at the time of the first detection is indicative of
the rapid luminosity rise near the time of shock breakout \citep{U07},
which is expected to occur about 1 day after the central explosion
for a red supergiant progenitor star.
We thus agree with the estimate of
\cite{li05} of an explosion date of 
September $22.0\pm 1.0$ (JD2453270.5).
The possible error in the date is of no consequence at the ages
of interest here.
 NGC 6946
is a nearby spiral galaxy \citep[5.5 Mpc,][]{sha01} that is largely
veiled at optical wavelengths by the Milky Way (Galactic longitude
$l=95^{\circ}$, latitude $b=+12^{\circ}$).   As revealed by
dust-penetrating near-infrared imaging, the galaxy is nearly face-on
($\sim30^{\circ}$  inclination) and its starlight extends over 12 arcmin
in diameter (see Figure \ref{SN2004etspitzerimage}), or $\sim$19 kpc in
physical size \citep{jar04, jar07}. The galaxy is gas-rich, actively
forming stars within giant molecular cloud complexes that trace the
multiple spiral-arm disk structure; the HI disk  extends well beyond the
starlight, doubling the size of the galaxy. NGC 6946 is a prolific
factory for supernova events.
Many supernova remnants (SNRs) 
have been identified using radio and X-ray telescopes
\citep{lac01, hym00, sch00, bla94, sch94, jar07}, and  SN 2004et is the
eighth historical SN found in NGC 6946.  Previous SNe include SN 2002hh \citep{li02}
and SN 1980K \citep{wil80}.

Optical spectroscopy of
SN 2004et showed it to be a Type II event 
  \citep{zwi04,fil04}, and the optical light curve
was that of a Type II plateau (IIP) supernova \citep{li05}.
\cite{sah06} found that 0.06$\pm0.02~ M_{\odot}$ of
$^{56}$Ni  and  $1.5-2 ~M_{\odot}$ of oxygen  were ejected in the
explosion, implying a progenitor mass of $\sim20 ~ M_{\odot}$.  
\cite{li05} identified a yellow supergiant progenitor candidate  using
presupernova optical images, implying a lower progenitor mass of
$15^{+5}_{-2} ~M_{\odot}$.  
Radio observations were carried by
\cite{sto04}, \cite{bes06}, and \cite{arg06}.  \cite{CFN06} modeled the
radio light curve, comparing with synchrotron and Compton
cooling models.  They estimated a mass loss rate of $\dot M_{-6}/u_{10}=
(9-10) T^{3/4}_{cs5}$, where $\dot{M}_{-6}$ is the mass loss rate in
units of $10^{-6}~M_{\odot}$ yr$^{-1}$, $u_{10}$ is the wind velocity in
units of $10$ km s$^{-1}$, and $T_{cs5}$ is the circumstellar
temperature in units of $10^5$ K. The {\it Spitzer Infrared Nearby
Galaxies Survey}  (SINGS; Kennicutt et al. 2003; see Figure
\ref{SN2004etspitzerimage}) imaged NGC 6946 after the 2004 supernova
event.  Using these data, a bright infrared source  was reported by
\cite{fab05}. \cite{RJC07} reported the detection of X-rays from a
position coincident with the infrared source, likely arising from the SN.

In this paper,  we give a more detailed presentation of the X-ray
emission from SN 2004et and implications of the X-ray emission for the
mass loss density. We discuss  the interaction model that is needed to
describe  the circumstellar interaction with ejecta in Type IIP
supernovae.

\section{OBSERVATIONS AND RESULTS}

\begin{figure}
\includegraphics[width=15truecm,angle=0]{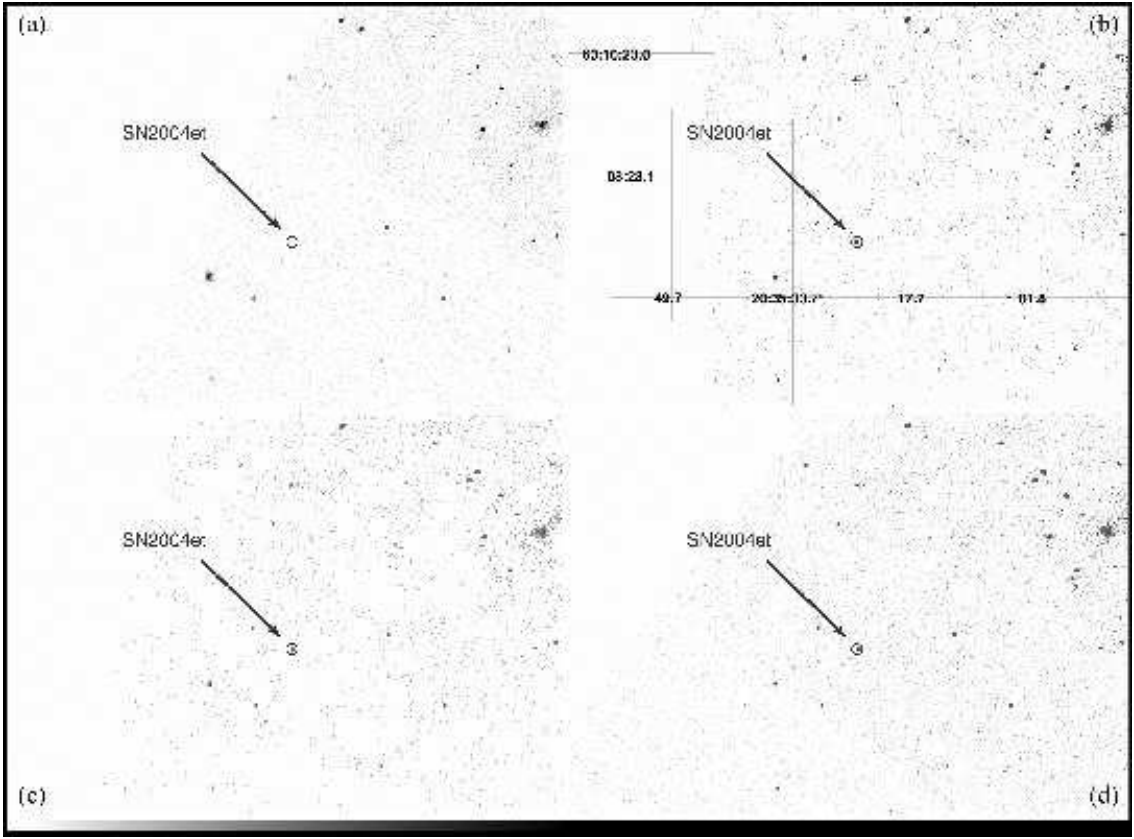}
\caption{X-ray {\it Chandra} images (0.4 - 8 keV) observed on 2002 November 25, 2002 (pre-SN; a), 2004 October 22 (b), 2004 November 6 (c), and 
2004 December 3 (d), showing detection of X-ray emission from SN 2004et (marked with a 
circle
of 5$''$ radius). 
}
  \label{SN2004etimage}
\end{figure}

Using archival {\it Chandra} data, we serendipitously identified X-ray
emission from SN 2004et. There are  two \chandra\ observations of NGC
6946 before the SN 2004et explosion: on 2001 September 7 and  2002
November 25  \citep{hol03}. No X-ray source is detected at the position
of SN 2004et using these two, pre-explosion observations (see Figure
\ref{SN2004etimage}).  Three \chandra\ observations (PI: Lewin) of NGC
6946 after the SN 2004et event were  carried out using the ACIS-S
detector on the \chandra\ {\it X-ray Observatory}.  SN 2004et lies
in the ACIS-S3 chip.     

\begin{figure}
\includegraphics[width=15truecm,angle=0]{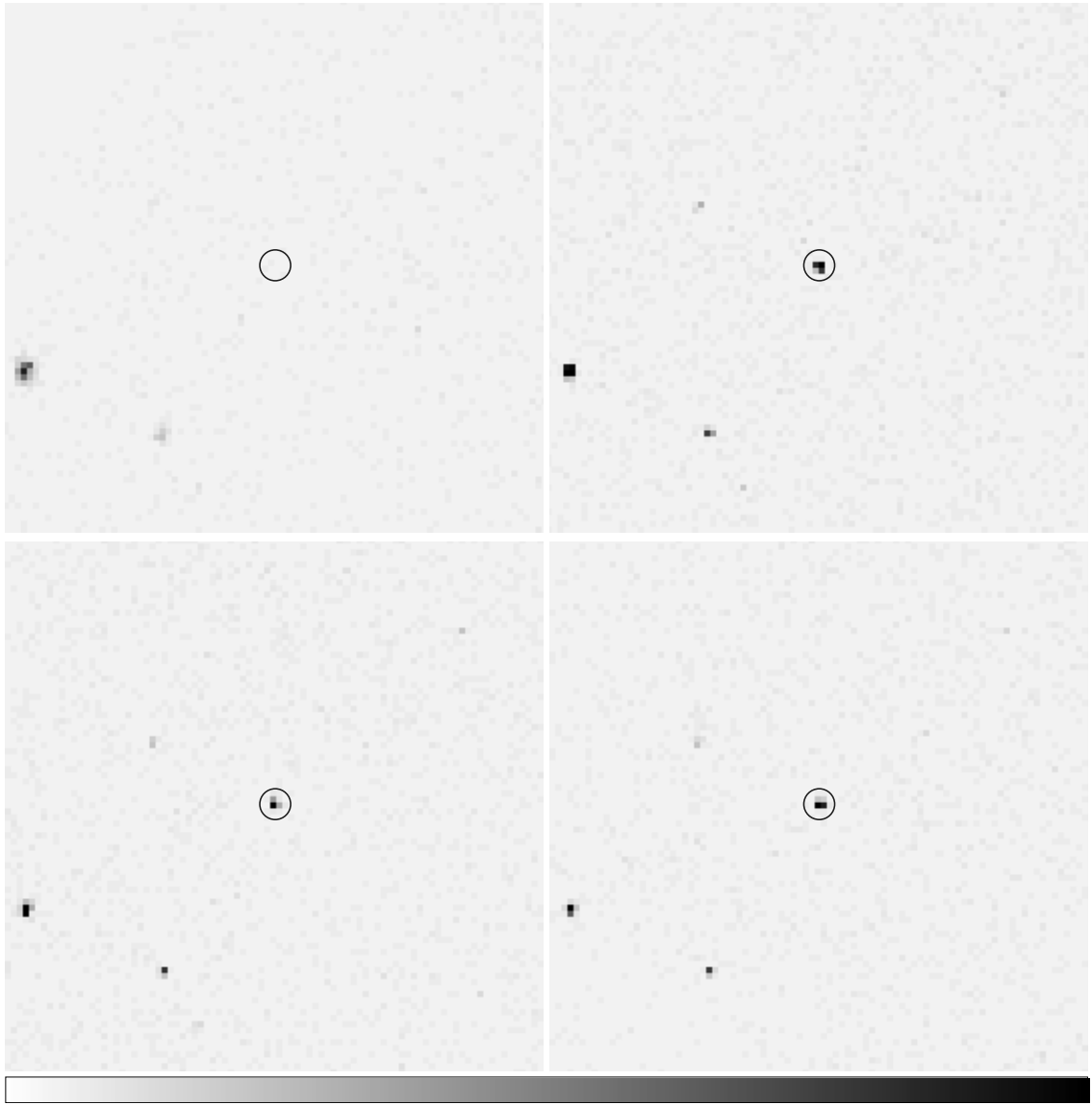}
\caption{Magnified version of Figure \ref{SN2004etimage} centered on SN 2004et. The image covers
a 3$'$ field of view. 
}
  \label{SN2004etimagezoom}
\end{figure}

We used all five sets of \chandra\ observations to study  the early  
evolution of SN 2004et.   We cross-correlated  the respective
positions of the USNO, 2MASS, and the previously published X-ray sources
\citep{hol03} with the positions  of our X-ray sources. There were a
few USNO/2MASS sources that coincided with X-ray  sources within
1$''$.  We found that the results of cross-correlated positions
conformed to our observations. An X-ray source appeared at the position
of SN  2004et for the three observations taken after the explosion.
  The  position of the X-ray source was found to agree with the
optical  position to  within $\sim$0.4$''$, consistent within the
position  uncertainty ($<$1$''$) for each of the three epochs. 
Figure \ref{SN2004etimage}  shows \chandra\ X-ray images of NGC 6946
covering SN 2004et before and  after the SN event.   Figure
~\ref{SN2004etimagezoom}, the magnified version centered on SN 2004et,
shows clear X-ray detection at the position of SN 2004et after the
explosion date.

For each epoch, we estimated the source counts using WAVDETECT (in
the   CIAO\footnote{http://cxc.harvard.edu/ciao/} package) with
the   reprocessed event data cube;  this tool performs  a Mexican hat
wavelet   decomposition and reconstruction of the image after
accounting for the   spatially varying point spread function as
described by \citet{fre02}.   We  extracted the counts for SN 2004et
using a 2 pixel ($\sim$2$''$)   circular aperture, enclosing $>$90\%
of the point-spread-function,   detecting 202, 151, and 158 photons
for the observations on 2004 October   22, 2004 November 6, and 2004
December 3, respectively. These dates   correspond to 30, 45, and 72
days after the explosion. In Table 1, we summerized the net counts,  
count rates and exposure times for three energy bands, $0.4-2$ keV,
$2-8$ keV   and $0.4-8$ keV. Throughout the paper, we constrain our
analysis to the energy bands  from 0.4 to 8 keV.   The X-ray  
source, SN 2004et, was not resolved in the \chandra\ image. We
estimated   a  hardness ratio (HR), defined as HR=(H$-$S)/(H+S) where
S is the soft   band ($0.4-2.0$ keV) and H is the hard band ($2.0-8$
keV), ranging from   $-0.237\pm0.070$ to $-0.454\pm0.085$ (see Table
1).  The hardness ratios   of the three observations indicate that the
spectra become softer   between 30 and 72 days after the explosion.

\begin{figure}
\includegraphics[width=8truecm,angle=270]{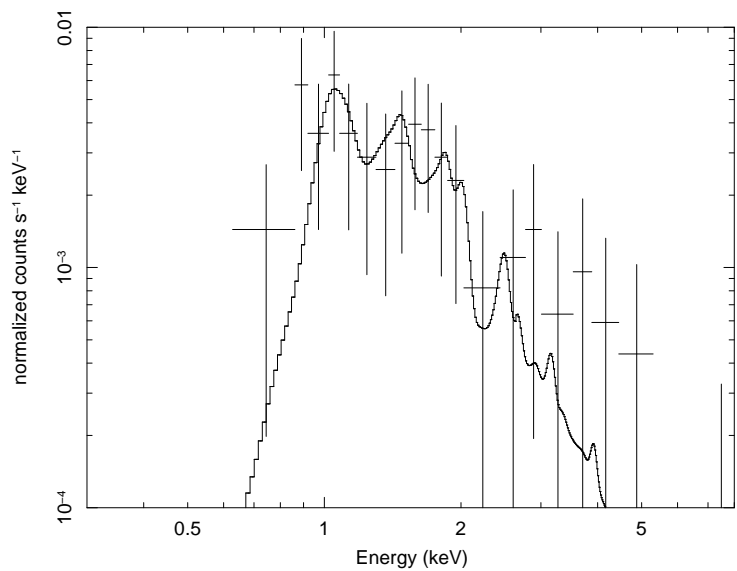}
\caption{X-ray spectrum of SN 2004et from the 2004 October 22 observation;
the best-fit model is superposed.
}
 \label{f-spec}
\end{figure}

\begin{figure}
\includegraphics[width=8truecm,angle=0]{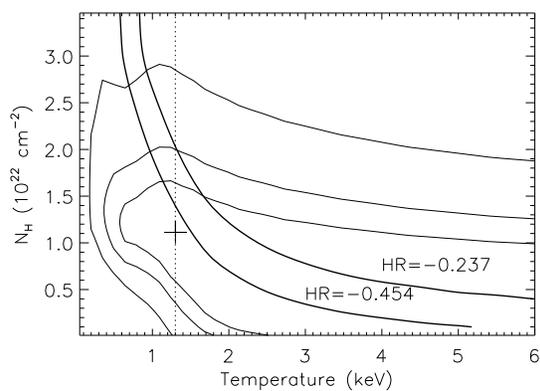} 
\caption{
 The confidence contours for column density and temperature solutions.
The confidence levels are 99\%, 90\% and 67\%.  The contours (thick lines) of the hardness
ratio of $-0.237$ and $-0.454$ are also shown.
For a given temperature, the softened X-ray spectrum between
the hardness ratio of $-0.237$ (the first) and of $-0.454$ (the last epoch)
indicates lowering in the absorption.  
}
 \label{f-conf}
\end{figure}

We also extracted spectra using PSEXTRACT in the CIAO package.  The  
relatively low number counts, $\sim$200 (2004 October 22
observation),   were barely sufficient to  obtain  a statistically
meaningful spectrum.   Accordingly we binned the spectra with a
minimum count of 10 per bin.   The  resulting spectrum of the 2004
October 22  observation, in Figure   \ref{f-spec}, is best fit by a
thermal model  with a temperature of   $kT=1.3^{+\infty}_{-0.8}$ keV
and a line-of-sight absorption of   $N_H=1.0\pm1.0 \times 10^{22}$
cm$^{-2}$. All errors associated with the spectral fitting results  
are within 90\% confidence.    The  column density and temperature
confidence contours are   also shown in Figure \ref{f-conf}. The
estimated extinction to the   supernova, $A_v$, is  0.41 mag
\citep{zwi04} using optical lines,   implying  a foreground atomic
hydrogen column density of $3 \times   10^{21}$ cm$^{-2}$ using
N$_H$/E(B-V) = 6.8$\pm$1.6$\times$10$^{21}$ H   atoms  cm$^{-2}$
mag$^{-1}$ (Gorenstein \& Tucker 1976). This extinction   is 
consistent with the lower envelope of our X-ray absorption  
measurements.   The thermal model uses  
APEC/MEKAL/RS\footnote{http://heasarc.gsfc.nasa.gov/}  models, which  
produced almost identical fit parameters to within 2\%.  The thermal
model which  includes  Ne X (at 1 keV), Mg XII (at 1.35 keV), and Si
XIII (at  1.865 keV) lines is consistent with the observed X-ray
spectrum, as shown in Figure \ref{f-spec}. The line emission  
indicates that the observed X-ray emission is largely dominated by  
reverse shock material at lower temperature than that of forward
shocked   material (see \S3 and \S4).    The inferred unabsorbed X-ray
luminosity   ($0.4-8$ keV)  is $3.8^{+1.8}_{-2.6} \times  10^{38}$ erg
s$^{-1}$,   using a distance of 5.5 Mpc, which falls  into the
luminosity range of   known X-ray SNe \citep{sch01, Imm03}.  Similar
consistency is found by fitting   a   bremsstrahlung model to the 2004
October 22 spectrum, yielding an   absorbing column density of
0.22$^{+0.78}_{-0.22} \times 10^{22}$ cm$^{-2}$,   and a temperature
of 25$^{+\infty}_{-24.5}$ keV.    The results of spectral fitting are
summarized in Table 2.

\begin{figure}
\includegraphics[width=8truecm,angle=270]{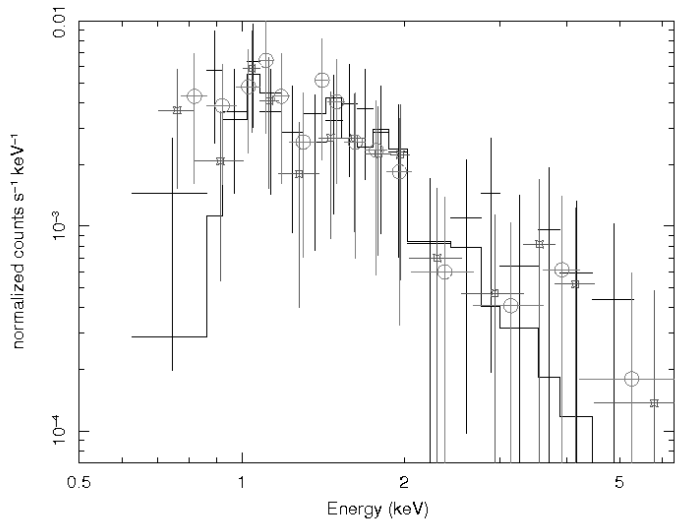}
\caption{{\it Chandra} spectra of SN 2004et at three epochs.
The spectra at the first, second, and third epochs
are marked with crosses, squares, and circles, respectively.
The spectral change below 1 keV is
 noticeable and the spectral
 softening might be due to decreasing shell absorption.
 The best-fit model of the first epoch spectrum
is superposed.} 
\label{threespec}
\end{figure}

Figure \ref{threespec} shows   comparisons of the spectra at the three
epochs.    The spectral fitting was not stable for the second and  
third observations, due to the diminishing number of X-ray photons, so
we   were unable to constrain the variation of temperature and
absorption by   using spectral fitting.   Assuming that the spectral
parameters were the same as the best-fit parameters  of the first
observations, and  using a thermal model (N$_H$=1.1$\times 10^{22}$
cm$^{-2}$   and kT= 1.3 keV),  we estimated fluxes and luminosities of
the second and third observations. The  corresponding fluxes were   
3.0$^{+0.5}_{-0.2}$$\times 10^{-14}$ and 3.4$^{+0.4}_{-0.3}$$\times
10^{-14}$    erg cm$^{-2}$ s$^{-1}$, respectively, and the
luminosities were 3.3$^{+0.4}_{-0.3}$$\times 10^{38}$ and  
3.6$^{+0.4}_{-0.3}$$\times 10^{38}$ erg s$^{-1}$, respectively;   
note that the errors of fluxes and luminosities were estimated  using
those of the count rates,  and the estimated former errors had lower
limits because   uncertainties of spectral fitting were not taken
account.   The respective luminosities and fluxes of the second and
third observations   were consistent with being the same as those of
the first observation.

\section{IMPLICATIONS FOR THE WIND DENSITY}

In the standard picture for the interaction of SN~IIP ejecta with a
smooth wind,  a double-shock structure forms  with the forward shock
propagating in the circumstellar (CS)  gas and the reverse shock in the
SN ejecta \citep{Che82a,Nad85}.  For the red supergiant  wind expected
in a SN~IIP, assuming $\dot{M}\sim 10^{-6}-10^{-5}~M_{\odot}$ 
\citep{CFN06}, the reverse shock dominates the X-ray luminosity. In the
case of a power law density distribution in the ejecta,  $\rho_{\rm
sn}\propto v^{-k}$, the interaction with the wind  in the self-similar
regime \citep{Che82b} gives a luminosity of the adiabatic reverse shock
at an age $t$ 
\begin{equation}
L=\xi(k-3)(k-4)^2\frac{w^{2}\Lambda N_{\rm A}^{2}}{4\pi v_{\rm sn}t}=
0.9\times10^{38}\omega^2\left(\frac{t}{30~\mbox{d}}\right)^{-1}
\left(\frac{v_{\rm sn}}{10^{9}\,\mbox{cm~s$^{-1}$}}\right)^{-1}\ergs \,,
\end{equation}

where $k=9$ is assumed,
$\xi=(1+X)(1+2X)/8=0.66$ for solar abundances,  $N_{\rm A}$ is
Avogadro's number, $v_{\rm sn}$ is the boundary (highest) ejecta 
velocity, and $w=\dot{M}/u$ is the wind density parameter, while
$\omega=\dot{M}_{-6}/u_{10}$ is the dimensionless  wind density
parameter,  where $\dot{M}_{-6}$ is the mass loss rate in units of
$10^{-6}~M_{\odot}$ yr$^{-1}$ and $u_{10}$ is the wind velocity in units
of $10$ km s$^{-1}$. The blue edge of H$\alpha$ in SN 2004et on day 30 
\citep{sah06} implies $v_{\rm sn}\approx12,500$ km s$^{-1}$. The
temperature at the reverse shock assuming equilibration  is $T_{\rm
r}=1.36\times 10^{9}(v_{\rm sn}/10^{9}  ~\mbox{km~s$^{-1}$})^{2}/(k-2)^{2}$
\citep{Che82b} and for the assumed parameters is equal to 3.7 keV.  The
cooling function for this temperature is  $\Lambda\approx
2\times10^{-23}$ erg s$^{-1}$ cm$^3$.  To produce the unabsorbed X-ray
luminosity of $\sim4\times10^{38}$ erg s$^{-1}$  on day 30, one then needs
a wind density $\omega\approx 2.5$. The pre-supernova mass loss rate 
of $(2-2.5)\times 10^{-6}~M_{\odot}$
yr$^{-1}$ for a wind velocity of 10 \kms.
The wind density is in line with estimates for other
Type IIP supernovae, such as SN 1999em \citep{poo02} and SN 2006bp \citep{Imm07}.

The temperature of the reverse shock predicted by the self-similar
model  (3.7 keV) is larger than the 1.3 keV  implied by the observed
spectrum, although it is within
the observational uncertainty.  
Incomplete temperature equilibration cannot be the reason for
this because equilibration is expected for the physical conditions that
are present. More plausible is that the velocity of the reverse shock is
lower due to  a deviation from  self-similar evolution. The formation of
a boundary shell during the shock breakout stage
\citep{GIN71,FA77,Che81,U07}  causes deviations from a power law density
distribution  in the outmost layers.  The effect is especially important
for SNe IIP because of their large progenitor radius and relatively low
peak velocities.

The existence of the boundary shell and  of the velocity cutoff at
$v_{\rm c}$  is implemented here in a numerical interaction model  based
on the thin shell approximation \citep{Che82b,CC06}. The mass of the
boundary shell is approximately equal to the mass of the outer envelope
($v>v_{\rm c}$) with an extrapolated power law $\rho\propto v^{-k}$. We
take the density distribution to be $\rho_{\rm sn}=\rho_0/(1+(v/v_0)^k)$
with  $k=9$ and $v_{\rm c}=15000$ km s$^{-1}$. The assumed ejecta mass
is $M=15~M_{\odot}$, which implies a main sequence  mass
$16.4-17~M_{\odot}$, in general agreement with   the observational
estimate of the progenitor mass, $15^{+5}_{-2}~M_{\odot}$  \citep{li05}.
The assumed kinetic energy is $E=1.3\times10^{51}$ erg, in line with a
model for the Type IIP SN~1999em \citep{U07}.

\begin{figure}
\plotone{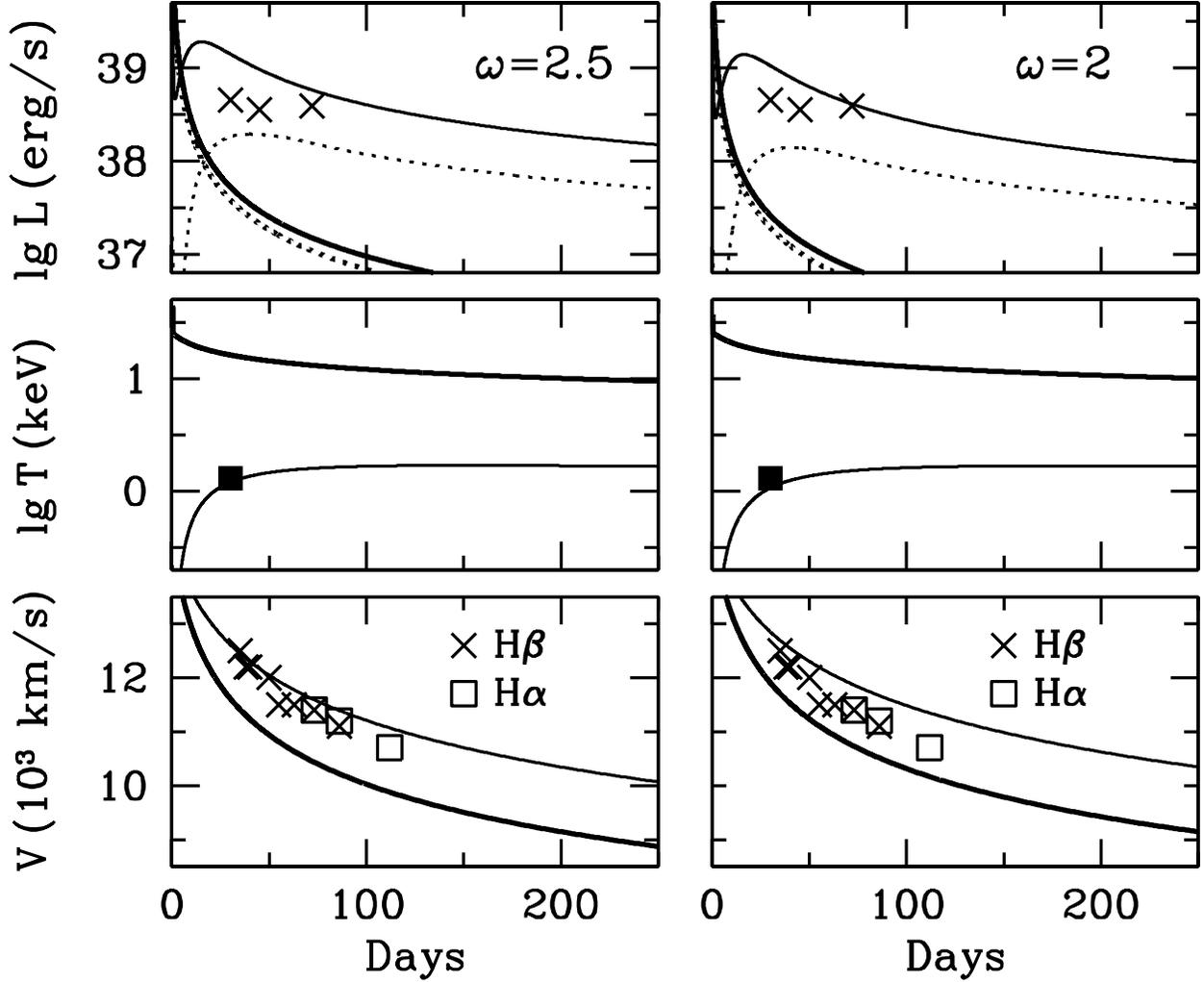}
\caption{The interaction model for two values of wind density $w=2.5$
and $w=2$. The {\em top} panel shows the X-ray luminosity of the forward
({\em thick} lines) and reverse ({\em thin}) shocks for unabsorbed ({\em
solid}) and absorbed ({\em dotted}) emergent  radiation.  The
absorbed radiation includes absorption  in the cool boundary shell (see
the text for details). The observed X-ray luminosities are shown by
{\em crosses}. The {\em middle} panel displays the electron temperature
of the forward ({\em thick} lines) and reverse ({\em thin}) shocks; the
observational estimate of the temperature, 1.3 keV on day 30,  is shown
by a filled {\em square}.  In the {\em lower} panel, the velocity of the
thin shell ({\em thick}  line) and boundary velocity of the unshocked SN
ejecta ({\em thin}  line)  are shown together with the velocities of the
blue  edge of the H$\beta$ and H$\alpha$ lines according to spectra of
\cite{sah06}.
 }
  \label{f-dyn}
  \end{figure}

Two expected effects related to the boundary shell are: 
(1)  non-self-similar early evolution of the shell deceleration that affects 
the velocity of the reverse shock and its X-ray emission; 
(2) significant absorption of X-rays in the cool boundary shell at the
reverse shock.
This contrasts with pure self-similar evolution, in which a  radiative
shell   does not typically form for the winds of interest, and  so the
absorption of reverse shock emission is small \citep{CFN06}.  In the
presence of a boundary shell with  mass $M_{\rm bs}$  the initial
velocity of the reverse shock  $v_{\rm rs}= v_{\rm sn}-v_{\rm s}$ is
about zero  once free expansion of the ejecta has been set up because
both velocities  coincide and are equal to the boundary velocity $v_{\rm
c}$. As the  shell decelerates, the velocity and  X-ray luminosity of
the reverse shock rapidly increase from  zero values (Figure \ref{f-dyn}).
The shell deceleration is surprisingly fast because only a small mass of
the swept up wind suffices to  reduce the shell velocity by  $\Delta
v\sim v_{\rm c}/(k-3)$, the value of the velocity jump between ejecta
and shell in the self-similar regime. For our model with  wind density
$\omega=2.5$,  $v_{\rm c}=15,000$ km s$^{-1}$, and mass of the boundary
shell  $M_{\rm bs}=4.7\times10^{-4}~M_{\odot}$,  the initial
deceleration takes $t_{\rm dec}\approx(M_{\rm bs}/wv_{\rm
c})(k-3)^{-1}\sim 8$ d. This does not mean that the self-similar regime
is completely  attained at the age  $t\sim t_{\rm dec}$.  Relatively
small deviations of the velocities  $v_{\rm sn}$ and $v_{\rm s}$ from
the self-similar values lead to a  significant deviation of $v_{\rm rs}$
from its self-similar value.  It takes at least  $t\approx(M_{\rm
bs}/wv_{\rm c})\sim(k-3)t_{\rm dec}$ or about 50 days  for the model of
interest to approach the self-similar expansion regime.  Computations
show that in the model with  a boundary shell, $v_{\rm rs}$ remains
lower than the self-similar value   by $>20\%$ during about first 200
days, primarily on account of the  lower ejecta velocity $v_{\rm sn}$
compared to the self-similar value.

The calculated X-ray luminosities (absorbed and unabsorbed)  of the
forward and reverse shocks, the electron temperatures in the shocks, and
velocities of the thin shell and highest velocity ejecta are shown  in
Figure \ref{f-dyn}  for two values of the wind density  parameter,
$\omega=2.5$ and 2.  The unabsorbed luminosity deduced from observations
is lower then the unabsorbed luminosity  in the interaction model
because the procedure of fitting the spectrum and the column density of
the absorber assumes a unique parameter $N_{\rm H}$. In reality the
emergent luminosity  is a combination of the radiation of the front and
rear sides of the reverse  shock and these components differ by the
amount of  absorption by the SN ejecta. On day 30, the optical depth at
1 keV  in the SN ejecta is $\tau\approx 15$, while all the remaining
components  (cool dense shell, CS wind and Galactic absorption) provide
only  $\tau\approx 3$.  Evidence for strongly absorbed  radiation from
the rear side of the reverse shock is seen for $E>2$ keV (Figure \ref{f-spec}).
The unabsorbed luminosity obtained from  X-ray
observations thus underestimates the real value by a factor 
$\approx2$.  Considering this, and taking into account  the uncertainty
of the distance, we conclude that the observed  X-ray emission from
SN~2004et implies a wind density $\omega\approx 2-2.5$. The electron
temperature of the reverse shock in the model with $\omega=2.5$ is very
close to the observational estimate (Figure \ref{f-dyn}), which  solves
the temperature problem in the self-similar model.  In our model, the
temperature at the reverse shock front remains fairly constant over the
time of interest (Figure \ref{f-dyn}), but the observed spectrum becomes
softer with time  because of  the decreasing column density through the
cool shell. This can be seen in the relative absorbed and unabsorbed
luminosities in Figure \ref{f-dyn}, and may be the reason for the spectral
softening seen in the observations.

A further test of the interaction model is provided by the  evolution of
the boundary velocity of the SN ejecta ($v_{\rm sn}$) and  the observed
blue edge velocity ($v_{\rm be}$) of  H$\alpha$ and H$\beta$ absorption
(Figure \ref{f-dyn}). Generally,  at the late photospheric stage, the
outer  layers of ejecta are considered to have recombined, so one
expects  that $v_{\rm be}$ should be significantly lower than $v_{\rm
sn}$. However, it was found  recently that ionization and excitation of
SN~IIP ejecta by X-rays emitted  from the reverse shock can produce
significant additional  excitation of hydrogen in the outer layers so
the  blue edge velocity $v_{\rm be}$ in H$\alpha$  can be very close to
$v_{\rm sn}$ \citep{CCU07}. Using the same procedure as in
\cite{CCU07},  we computed the excitation  of hydrogen for a model of
SN~2004et ejecta with X-ray irradiation  effects for $\omega=2$ and
$\omega=2.5$. In both cases,  the additional high velocity absorption in
H$\alpha$ and H$\beta$  turns out to be strong and the blue edge of
absorption is only $\sim250$ km s$^{-1}$ lower than the boundary
velocity $v_{\rm sn}$.  The lower panel of Figure \ref{f-dyn}  shows that
the model with $\omega=2.5$ successfully passes   the velocity test and
therefore is preferred to the $\omega=2$ model. For $\omega=2$, the
interaction model predicts somewhat larger values of $v_{\rm sn}$.
However, given the uncertainty in the choice of the  initial velocity
cut-off $v_{\rm c}$, this test should not be considered  as conclusive,
although it demonstrates the self-consistency of the model.

\section{DISCUSSION}

We detect the X-ray emission from SN 2004et using the 
{\it Chandra X-Ray Observatory} data and interpret it in terms of the 
interaction of ejecta with a presupernova wind characterized by the 
wind density parameter $\omega=2-2.5$ or 
$\dot{M}=(2-2.5)\times10^{-6}~M_{\odot}$ yr$^{-1}$ 
assuming the wind velocity of 10 \kms.
It should be kept in mind that the derived wind density 
is model dependent and also suffers from the error 
in the observed X-ray luminosity. Both result in an 
uncertainty in the wind density of at least 0.3 dex.
Interestingly, the  wind density deduced for SN~2004et is a factor 
of two larger than the wind density for SN~1999em and SN~2004dj 
with $\omega\approx 1$ as estimated from optical effects of the ejecta-wind 
interaction \citep{CCU07}. This fact 
possibly reflects a real difference, because the radio 
luminosity of SN 2004et is among highest for SNe~IIP \citep{CFN06}.

The preferred wind density in SN~2004et ($\omega=2.5$) is consistent 
with the value found from free-free absorption of radio emission 
\citep{CFN06}, if the assumed temperature of the wind is  $T_{\rm
w}\approx15,000$ K. We computed the wind temperature irradiated by the
X-rays from the  reverse shock in the steady state approximation (barely
applicable  at the relevant age of $30-50$ d) and found $T_{\rm
w}\approx13,000$ K  in the case of $\omega=2.5$. This makes the radio
and X-ray properties of the wind in SN~2004et consistent with a wind
density parameter  $\omega=2.5$. The value of $\dot M$ deduced for SN
2004et is consistent with the mass loss rate expected for a progenitor
star of its mass, $(15-20)~M_{\odot}$ \citep[Figure 1 of][]{CFN06}. We
have assumed that SN 2004et had a normal red supergiant progenitor, but
\cite{li05} found that the progenitor might be a yellow supergiant, with
possible similarities to the progenitors of SN 1987A or SN 1993J.
However, its radio and X-ray properties are similar to those of other
Type IIP supernovae \citep[][and references therein]{CFN06} and unlike
those of SN 1987A and SN 1993J. Also, the optical light curve of SN
2004et is typical of a SN IIP \citep{sah06}, which would seem to require
a red supergiant progenitor. 
Moreover, as noted above, the wind density around SN~2004et is
relatively large for a SN~IIP.
Yet for the same stellar mass a yellow supergiant is expected 
to have a similar mass loss rate but a larger wind velocity, and, 
therefore, a lower wind density compared to the red supergiant,  
in contrast with the estimated wind density in SN~2004et. This
additionally argues in favor of the identification of the 
SN~2004et progenitor with a red supergiant.
We conjecture that the ground-based
observations of the progenitor combine the emission from a red
supergiant with emission from an earlier type star or stars.


\acknowledgements
We are grateful to the referee for a very helpful report.
This research was supported in part by NSF grant AST-0307366 (NNC and
RAC). J. R.  and T. H. J. acknowledge the support of California
Institute of Technology, the Jet Propulsion Laboratory, which is
operated under contract with NASA. 
This work is based on {\it Chandra}
archival data.

{}

\clearpage
\begin{deluxetable}{llllcccccclll}
\tabletypesize{\footnotesize}
\rotate
\tablenum{1}
\tablewidth{0pt}
\tablecaption{Summary of observations}
\label{chaobs1}
\tablehead{
\colhead{date } &\colhead{OBSID$^a$} &\colhead{net counts }& \colhead{snr$^b$} & \colhead{exposure} 
& \colhead{hard counts$^c$}  &  \colhead{soft counts$^c$}   & \colhead{HR$^c$} \\
\colhead{(days$^d$) } &\colhead{ } &\colhead{ } & \colhead{} & \colhead{time (ks)} 
& \colhead{}  &  \colhead{}  &  \colhead{}   
}
\startdata
2004 October 22  (30) & 4631 &201.8$\pm$16.7  &  12.8 & 29.7   & 133.5$\pm$11.5&82.4$\pm$9.11&$-0.237\pm$0.070(38\%) \\
2004 November 06 (45) &4632 &150.7$\pm$14.6  &  10.3 & 28.0 & 105.6$\pm$10.3&53.5$\pm$7.3 &$-0.327\pm$0.084(34\%) \\
2004 December 03 (72) &4633&158.3$\pm$14.7  &  10.8 & 26.6  & 121.6$\pm$11.0&45.6$\pm$6.8 &$-0.454\pm$0.085(27\%) \\
\enddata
\tablenotetext{a}{observation id. $^b$snr: signal-to-noise.
$^c$The  hardness ratio (HR) is defined as HR=(H$-$S)/(H+S) where S is the soft
band ($0.4-2.0$ keV) and H is the hard band ($2.0-8$ keV). The number in the 
parenthesis is percent of hard band photons. 
$^d$days since the explosion.
}
\end{deluxetable}

\begin{deluxetable}{ccllcccc}
\tabletypesize{\scriptsize}
\tablenum{2}
\tablewidth{0pt}
\tablecaption{Results of spectral fitting using the first epoch observation}
\label{chaobs2}
\tablehead{
\colhead{Model } & \colhead{$^a\Delta \chi2$} & \colhead{N$_H$}  & \colhead{kT}& \colhead{Z/Z$_{\odot}$}& \colhead{f$_{\rm x}$$^b$} & \colhead{L$_{\rm x}$$^c$}\\
\colhead{} & \colhead{} & \colhead{(10$^{22}$cm$^{-2}$)}  & \colhead{ (keV)}& \colhead{}& \colhead{(erg cm$^{-2}$ s$^{-1}$)} & \colhead{(erg s$^{-1}$)}\\
}
\startdata
thermal& 0.4$^c$  &1.1($<$2.1)  & 1.3 ($>$0.4)& $\equiv$1 &  2.9$^{+4.7}_{-1.4} \times 10^{-14}$&  3.8$^{+1.8}_{-2.6}$$\times 10^{38}$  \\
thermal& 0.18 & $\equiv$0.3   &     8.9($>$1.5) &$\equiv$1 &7.1($>$4.3)$\times 10^{-14}$   &   3.2($>$2.4)$\times 10^{38}$  \\
thermal$^d$& 0.2  &0.23 ($<$1.3) & 22 ($>$0) &0.2 & 6.6$\times 10^{-14}$&   
2.8$\times 10^{38}$\\
bremstrahlung&0.18   &0.22 ($<$1) & 25 ($>$0.5) &--& 7.5$^{+2.2}_{-5.8}$$\times 10^{-14}$&   3.3$^{+2.4}_{-1.2}$$\times 10^{38}$ \\
\enddata
\tablenotetext{a}{the typical degree of freedom is $\sim$16. $^b$absorbed flux, $^c$unabsorbed luminosity (0.4-8 keV).
$^d$ the errors of the abundance, flux and luminosity could not be constrainted.
 }
\end{deluxetable}

\clearpage


\end{document}